# Fast response photogating in monolayer MoS$_2$ phototransistors

Daniel Vaquero,[a] Vito Clericò[a], Juan Salvador-Sánchez[a], Elena Díaz[b], Francisco Domínguez-Adame[b], Leonor Chico[b,c], Yahya M. Meziani[a], Enrique Diez[a] and Jorge Quereda[a]*

[a.] *Nanotechnology Group, USAL–Nanolab, Universidad de Salamanca, E-37008 Salamanca, Spain.*
[b.] *GISC, Departamento de Física de Materiales, Universidad Complutense, E-28040 Madrid, Spain.*
[c.]*Instituto de Ciencia de Materiales de Madrid, CSIC, E-28049 Madrid, Spain.*
*J.Quereda@usal.es

**Abstract**: Two-dimensional transition metal dichalcogenide (TMD) phototransistors have been object of intensive research during the last years due to their potential for photodetection. Photoresponse in these devices is typically caused by a combination of two physical mechanisms: photoconductive effect (PCE) and photogating effect (PGE). In earlier literature for monolayer (1L) MoS$_2$ phototransistors PGE is generally attributed to charge trapping by polar molecules adsorbed to the semiconductor channel, giving rise to a very slow photoresponse. Thus, the photoresponse of 1L-MoS$_2$ phototransistors at high-frequency light modulation is assigned to PCE alone. Here we investigate the photoresponse of a fully h-BN encapsulated monolayer (1L) MoS$_2$ phototransistor. In contrast with previous understanding, we identify a rapidly-responding PGE mechanism that becomes the dominant contribution to photoresponse under high-frequency light modulation. Using a Hornbeck–Haynes model for the photocarrier dynamics, we fit the illumination power dependence of this PGE and estimate the energy level of the involved traps. The resulting energies are compatible with shallow traps in MoS$_2$ caused by the presence of sulfur vacancies.

## Introduction

Two-dimensional (2D) transition metal dichalcogenides (TMDs) are very attractive for the development of phototransistors and other optoelectronic devices at the nanoscale[1–5] due to their optical bandgap spanning the visible spectrum, large photoresponse, and high carrier mobility. In 2D TMD phototransistors, photoresponse typically stems from two main mechanisms:[6–12] The photoconductive effect (PCE), where light-induced formation of electron–hole pairs leads to an increased charge carrier density and electrical conductivity; and the photogating effect (PGE),[9] where the light-induced filling or depletion of localized states causes a shift of the Fermi energy. When the characteristic relaxation times for these localized states are very long, the light-induced Fermi energy shift persists long time after exposure to light. In this case, the effect is commonly referred as photodoping.[13,14]

The occurrence of PGE in 2D-TMD phototransistors is usually associated to the presence of polar molecules adsorbed onto the monolayer surface,[6] resulting in a very slow, atmosphere-dependent photoresponse. Thus, the general understanding is that PGEs can be ruled out simply by modulating the intensity of the optical excitation at relatively fast frequencies (~10 Hz). The high-frequency response of the device is therefore generally attributed to PCE.

Here, we investigate the photoresponse of a high-quality h-BN encapsulated monolayer MoS$_2$ phototransistor. In stark contrast with previous understanding, the dependence of the observed photoresponse on the gate voltage and illumination power indicates that PGE is the dominant contribution to photoresponse, even for light-modulation frequencies of up to 1 kHz, much faster than the response time of PGEs described in earlier literature.[6] Further, the observed fast-responding PGE remains present even when measuring at cryogenic conditions, where the characteristic times for charge trapping processes involving adsorbed polar molecules should be very long. This suggests the presence of an additional contribution to PGE, not related to adsorption of environmental species but instead caused by impurities in the MoS$_2$ crystal lattice.

The contribution to photoresponse coming from PGE only fades away when the semiconductor channel is in its off state, *i.e.,* for gate voltages $V_g$ well below the threshold voltage $V_{th}$. In this regime, the remaining photoresponse becomes linear with the illumination power, as expected for PCE.

We analyze the dynamics of photoexcited carriers using a Hornbeck–Haynes model[6,15] that accounts for PGEs caused by charge trapping at shallow impurities in the MoS$_2$ monolayer (not considered in previous works[6]). The model allows us to fit with great accuracy the experimentally observed power dependence of photocurrent and extract values for the density of localized states and the characteristic times for filling and depletion of charge traps. Finally, by considering the detailed balance principle, we estimate that the localized states involved in photogating lay at an energy ~ 8.4 meV above the valence-band edge. This estimated energy is compatible with shallow trap-states associated to sulfur vacancies, generally present in 2D-MoS$_2$.[16–18] Thus, our results suggest that the dominant mechanism for high-frequency photoresponse in monolayer MoS$_2$ phototransistors is a sulfur vacancy-mediated PGE, and not PCE as generally assumed in earlier literature.

## Results and discussion

**Photoconductive and photogating effects**

The inset in Figure 1a schematically shows the 1L-MoS$_2$ transistor geometry: The semiconductor channel is encapsulated between multilayer hexagonal boron nitride (h-BN) flakes in order to better preserve its intrinsic properties[19] and Ti/Au electrodes are fabricated on top following an edge-contact geometry (further described in the Methods section). The device is fabricated on a 300 nm SiO$_2$/Si substrate and the bottom Si layer is used as back gate. All the measurements reported in the main text are performed in vacuum and at $T$ = 5 K unless otherwise specified. Similar measurements at room temperature can be found in the Electronic Supplementary Information, section S1.

Figure 1a shows two-terminal $I$-$V$ curves of the monolayer MoS$_2$ phototransistor, measured both in the dark and while exposing the entire area of the device to uniform illumination with power density $P_D = 1$ mW mm$^{-2}$ and photon energy $h\nu = 1.92$ eV (on resonance with the X$^A$ exciton transition of 1L-MoS$_2$). The $I$-$V$ curves present a back-to-back diode-like behaviour due to the presence of Schottky barriers at the contacts.[20,21] The different saturation currents for positive and negative voltages are caused by an asymmetry in the Schottky barrier heights. Upon illumination, the drain-source current $I_{DS}$ increases by $I_{PC}$ due to PCE and PGE. The light-induced increase of current, $I_{PC}$, can be written as

$$I_{PC} = \Delta I_{PCE} + \Delta V_{PGE}\frac{dI_{ds}}{dV_g},\quad (1)$$

where $\Delta I_{PCE}$ is the increase of $I_{DS}$ caused by PCE, and $\Delta V_{PGE}$ is the effective change in the gate threshold voltage caused by PGE. It is worth noting that, at $V_{ds} = 0$ the photocurrent fades away, indicating that photovoltaic effects (which may occur at the metal/MoS$_2$ interfaces) do not give a measurable contribution to $I_{PC}$ for our experimental configuration.

Figure 1b shows gate transfer characteristics of the device acquired in the dark and under illumination. In the following, the drain-source voltage is kept at $V_{ds}$ = 10 V for consistency. However, the results presented below for the dependence of $I_{PC}$ on the gate voltage, illumination power and light modulation frequency do not change significantly for lower $V_{sd}$.

At low temperature, the transfer curves are almost hysteresis-free, showing a clear n-type behaviour, and the semiconductor channel conductivity increases as the back-gate voltage $V_g$ becomes larger than the threshold voltage $V_{th}$. The two contributions to $I_{PC}$ from equation 1 can be clearly distinguished in Figure 1b. There, the effect of PGE is observed as a horizontal shift of the transfer curve upon illumination, by the amount $\Delta V_{PGE}$, while PCE results in a smaller but measurable vertical shift by $\Delta I_{PCE}$ (see inset in the figure).

The increase in photocurrent caused by the PCE is given by

$$\Delta I_{PCE} = \frac{W}{L}V_{ds}\Delta\sigma_{PCE},\quad (2)$$

where W/L is the aspect ratio of the semiconductor channel, $V_{ds}$ is the drain-source voltage and $\Delta\sigma_{PCE}$ is the light-induced increase in conductivity due to the optically excited charge carriers:

$$\Delta\sigma_{PCE} = q(\mu_n n_{ph} + \mu_p p_{ph}).\quad (3)$$

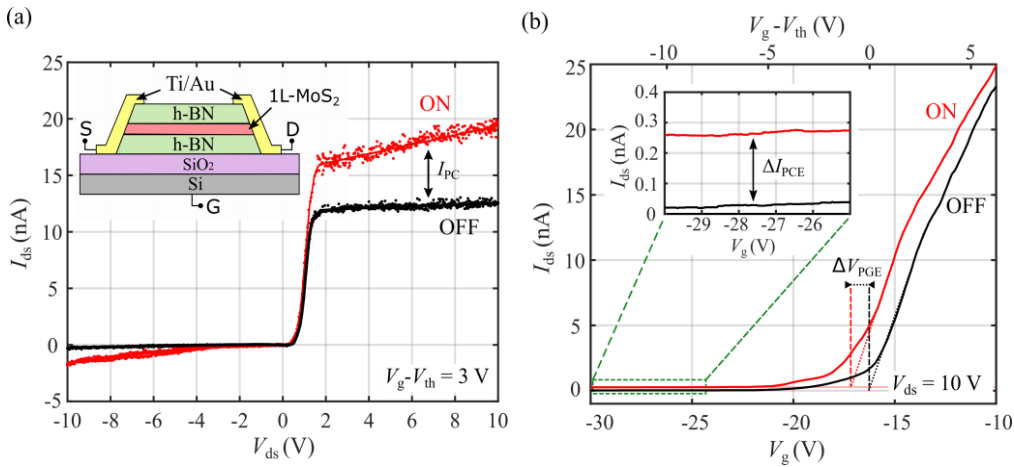

**Figure 1** – Electrical and optoelectronic response of the monolayer MoS$_2$ phototransistor. (a) Two-terminal $I$-$V$ characteristic of the monolayer MoS$_2$ phototransistor in the dark and under uniform illumination with power density $P_D = 1$ mW mm$^{-2}$ and photon energy $h\nu = 1.92$ eV. Upon illumination the drain-source current, $I_{ds}$ increases by $I_{PC}$. Inset: Schematic drawing of the device. (b) Gate transfer curves of the device, showing a threshold gate voltage $V_{th} = -11$ V. The inset shows a zoom-in of the region indicated by the dashed green rectangle. The contributions to photoresponse by $\Delta I_{PCE}$ and $\Delta V_{PGE}$ (see equation 1) are indicated in the plot.

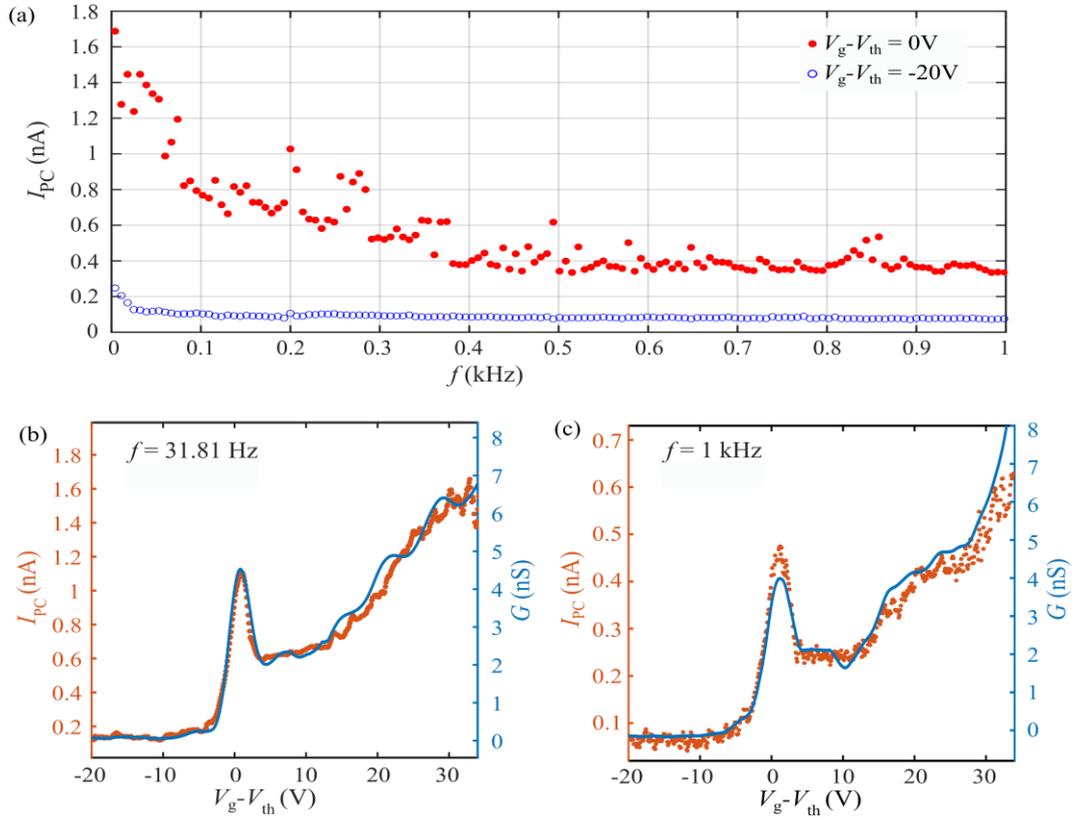

**Figure 2** – Frequency-dependent photoresponse. (a) $I_{PC}$ as a function of the light-modulation frequency. Measurements are shown for $V_g - V_{th} = -20$ V (blue, empty circles; corresponding to the PCE-dominated regime) and for $V_g - V_{th} = 0$ V (red, filled circles; PGE-dominated regime). (b) Transconductance (blue line, right axis) and gate-dependent photocurrent (orange dots, left axis) measured at $V_{ds} = 10$ V for illumination on resonance with the $X^A_{1s}$ exciton transition and a light-modulation frequency $f = 31.81$ Hz. (c) Same as (b) with $f = 1$ kHz.

Here, $\mu_n$ and $\mu_p$ are the electron and hole mobilities respectively, and $n_{ph}$ and $p_{ph}$ are the densities of optically generated excess charge carriers.

As discussed above, PGE appears when optically excited carriers can fall into trap states. While these trapped carriers do not directly contribute to transport, their presence can result in a partial screening of the gate voltage $V_g$, modifying the effective threshold voltage $V_{th}$ of the device, and consequently, the measured current. Assuming that in equilibrium there is a finite density $n_t$ of trapped carriers, we can use a parallel-plate capacitor model to estimate the shift $\Delta V_{PGE}$:

$$\Delta V_{PGE} = \frac{n_t e}{C_{ox}}, \quad (4)$$

where $e$ is the elementary charge and $C_{ox}$ is the capacitance of the h-BN/SiO$_2$ insulating layer. The resulting photocurrent $I_{PGE}$ is given by

$$\Delta V_{PGE} = \frac{n_t e}{C_{ox}}, \quad (5)$$

$$I_{PGE} = \frac{n_t e}{C_{ox}} \frac{dI_{ds}}{dV_g}. \quad (6)$$

Thus, $I_{PGE}$ is proportional to the transverse conductance $dI_{ds}/dV_g$, which enables us to distinguish it from $I_{PCE}$, as discussed below.

**Frequency dependence of $I_{PC}$**

We now consider the effect of the light-modulation frequency in the 1L-MoS$_2$ photoresponse. At this point it is useful to compare our results with a previous characterization of photoresponse in a monolayer MoS$_2$ phototransistor, reported by Furchi et al.[6] There, while measuring at room temperature, they observed a slow-responding PGE, which they attributed to charge-trapping by few layers of surface-bound water molecules underneath the MoS$_2$ sheet. By using a mechanical chopper to modulate the optical excitation and registering the signal with a lock-in amplifier, they observed that the photocurrent $I_{PC}$ largely decreased for light-modulation frequencies above ~1 Hz, as the trapping process was too slow to respond to the excitation. Thus they interpreted the remaining high-frequency signal as originated by PCE.

For comparison, we now also make use of a lock-in amplifier to measure the dependence of $I_{PC}$ on the light modulation frequency for our device, as shown in Figure 2a. Similarly to Furchi et al. we also observe a reduction of the signal at higher frequencies. For our measurements at $T$ = 5 K, we find that $I_{PC}$ decreases by roughly a factor 3, while at room temperature we observe a much larger reduction (see Supplementary Section S1). This weaker reduction at cryogenic temperatures is compatible with the slow-responding PGE caused by adsorbed polar molecules, since the effect of these dipoles should largely decrease at cryogenic temperatures. Interestingly, we find that the frequency dependence of the signal can be modified with the gate voltage, with $I_{PC}$ decaying much more slowly with the modulation frequency for larger gate voltages.

Let us now investigate the origin of the remaining signal for high-frequency modulation. As mentioned above, this fast-response contribution to the photocurrent is usually attributed to PCE in earlier literature. However, as we argue below, we find that the behaviour of this fast-response photocurrent can be better described by considering an additional contribution to PGE.

A characteristic signature of PGE is that the resulting photocurrent $I_{PC}$ is proportional to the transconductance $G = dI_{ds}/dV_g$ of the semiconductor channel (see equation 6). This allows us to clearly distinguish it from PCE, which should not have a strong dependence on $V_g$ for low carrier densities. As shown in Figure 2b, we find that for our 1L-MoS$_2$ device the $V_g$-dependence of $I_{PC}$ is very strongly correlated to the transconductance $G$ (obtained as the numerical derivative of the $I$-$V$ transfer characteristic). Importantly, this remains true even when the light is modulated at frequencies as high as 1 kHz (Figure 2c). This trend indicates that the photoresponse is mainly dominated by PGE even at high frequency, in stark contrast with earlier understanding.[6] As discussed below, we attribute this fast-response PGE to charge trapping at sulfur vacancies, present in the 1L-MoS$_2$ crystal.

At gate voltages well below $V_{th}$ the device shows a smaller, but measurable photocurrent. In this regime the transconductance $G$ is zero and, consequently, the PGE contribution to $I_{PC}$ fades away. We conclude that the small remaining photocurrent for $V_g \ll V_{th}$ must be caused by PCE.

**Power density dependence of $I_{PC}$**

To further confirm our interpretation of the photoresponse for the two gate voltage regimes ($V_g \gg V_{th}$ and $V_g \ll V_{th}$) we now study the dependence of $I_{PC}$ on the illumination power density. Figure 3a shows $I_{PC}$ as a function of the illumination power for $V_g - V_{th} = -20$ V at two different photon energies, corresponding to the $X_{1s}^A$ (1.92 eV) and $X_{1s}^B$ (2.07 eV) excitonic transitions of 1L-MoS$_2$ (see Supplementary Section S2). In both cases, $I_{PC}$ increases linearly with the power density, $P_D$. As we discuss in the section below, this is the expected power dependence of $I_{PC}$ for pure PCE.

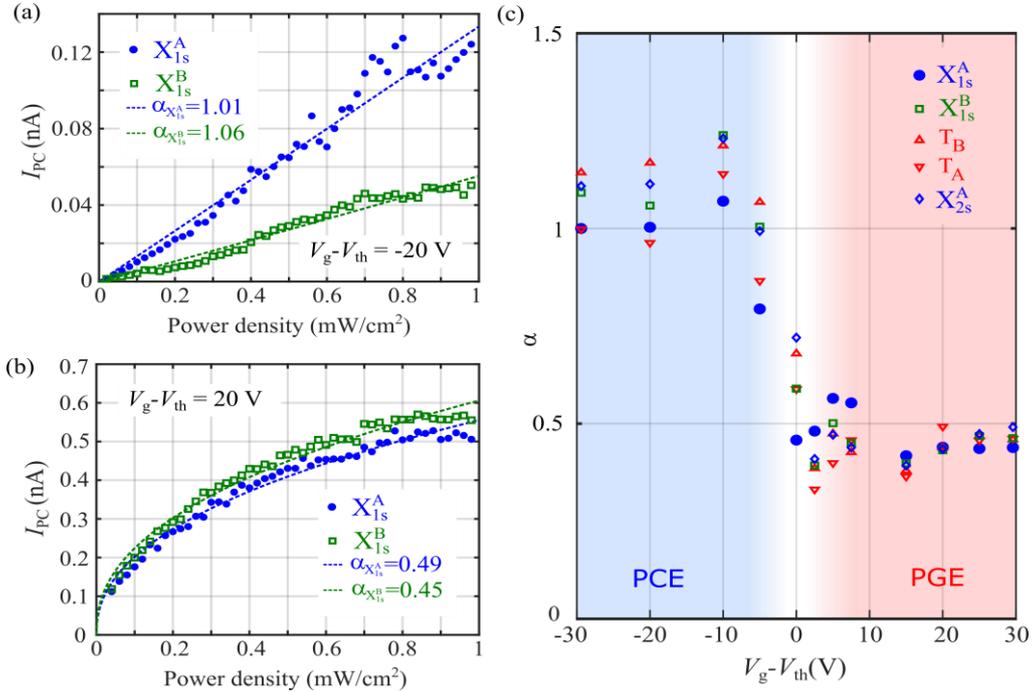

**Figure 3** – Power dependence of $I_{PC}$ in the two gate voltage regimes. (a) Power dependence of $I_{PC}$ for $V_g - V_{th} = -20$ V acquired at two different photon energies $E$ matching the $X_{1s}^A$ and $X_{1s}^B$ excitonic spectral features of monolayer MoS$_2$. Lines are fittings to $I_{PC} \propto P_D^\alpha$ with $\alpha \approx 1$. (b) Same as (a) for $V_g - V_{th} = 20$ V. The fittings now give $\alpha \approx 0.5$. (c) Dependence of the fitting parameter $\alpha$ on $V_g - V_{th}$, measured for photon energies matching the five main excitonic spectral features of monolayer MoS$_2$. The two different regimes for power dependence, corresponding to the PCE-dominated and the PGE-dominated photoresponse regimes are indicated in the figure.

For gate voltages well above the threshold voltage (Figure 3b), however, the situation completely changes and the power dependence of $I_{PC}$ becomes sublinear. A typical phenomenological approach used in previous works to distinguish PGE and PCE is to fit the power dependence to $I_{PC} \propto P_D^\alpha$, where $\alpha = 1$ is generally associated to PCE and $\alpha < 1$ to PGE. Figure 3c shows the parameter $\alpha$ extracted from these fittings as a function of the gate voltage for five different illumination energies, matching the five main excitonic transitions of 1L-MoS$_2$, as labelled in the figure and discussed in Supplementary Section S2. As one can clearly observe in the figure, for gate voltages below $V_{th}$ we get $\alpha \approx 1$, regardless of the selected illumination wavelength, while for $V_g > V_{th}$ we get $\alpha \approx 0.5$. In the next section we discuss the photocarrier dynamics of the system and correlate them with the observed power dependencies.

**Carrier dynamics**

Proceeding similarly to earlier literature[6,15] we analyse the dynamics of photoexcited carriers using a modified Hornbeck–Haynes model. We consider a scenario where the main photocarrier relaxation mechanism is Shockley-Read-Hall recombination mediated by midgap states. We also include a discrete density of localized states $D_t$ at an energy near the valence band edge to account for the presence of shallow hole traps (See Figure 4a). In 1L-MoS$_2$ such midgap states and shallow traps are expected to occur due to the presence of sulfur vacancies in the crystal lattice.[16–18] For an n-doped semiconductor we can assume that only the hole traps near the valence band are relevant, since electron traps are already filled at equilibrium. For simplicity, we also assume that the characteristic times for decay of electrons and holes to the midgap states are equal, i.e., $\tau_e = \tau_h \equiv \tau_r$.[6] Under these assumptions, the dynamics of the photoexcited carriers are described by:

$$\frac{dn_{ph}}{dt} = \phi_A - n_{ph}\tau_r^{-1}, \tag{7}$$

$$\frac{dp_{ph}}{dt} = \phi_A - p_{ph}\tau_r^{-1} - p_{ph}\tau_t^{-1}\left(1 - \frac{p_t}{D_t}\right) + p_t\tau_d^{-1}, \tag{8}$$

$$\frac{dp_t}{dt} = p_{ph}\tau_t^{-1}\left(1 - \frac{p_t}{D_t}\right) - p_t\tau_d^{-1}. \tag{9}$$

Here, $D_t$ is the density of localized states, $p_t$ is the density of trapped holes, and $\tau_t$ and $\tau_d$ are the characteristic times for trapping and detrapping of holes into these states, respectively. $\phi_A$ is the density of absorbed photons, related with the power density by $\phi_A = \eta P_D \lambda/hc$, being $\eta$ the optical absorption of MoS$_2$ and $\lambda$ is the illumination wavelength.

Solving equations (7-9) for the steady state we get:

$$p_{ph} = \phi_A \tau_r, \tag{10}$$

$$p_t = \frac{\phi_A D_t \tau_r}{\phi_A \tau_r + D_t \left(\frac{\tau_t}{\tau_d}\right)}. \tag{11}$$

The presence of hole traps has two main effects in the resulting photoresponse: Firstly, it affects the efficiency of PCE relative to

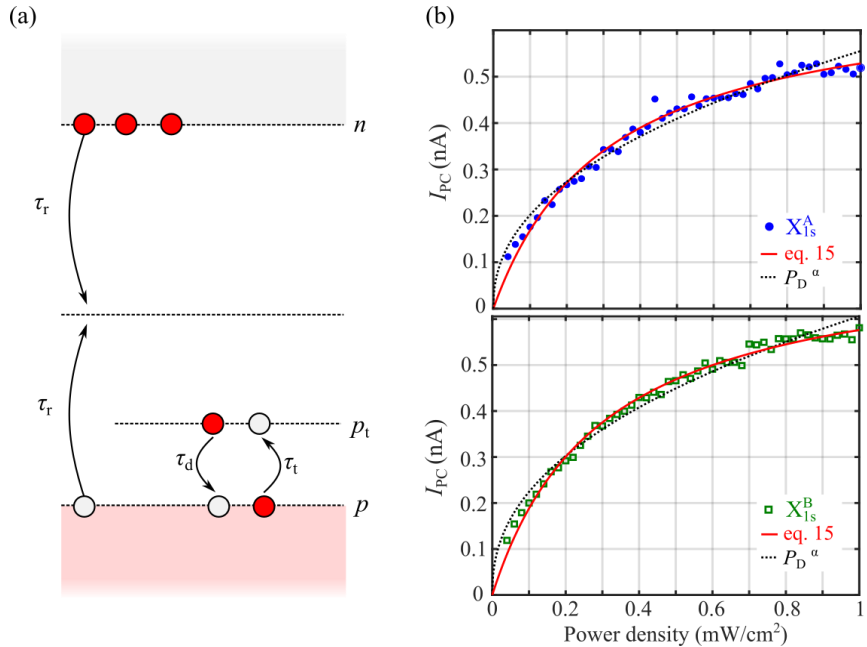

**Figure 4** – Model for photocarrier dynamics. (a) Schematic drawing of the simplified energy band diagram and excitation/relaxation processes considered in the model. (b) Comparison between fittings of the power dependence of $I_{PC}$ to $P_D^\alpha$ (black dashed lines) and to our model (equation 15; red solid lines)

a trap-free scenario. The photoinduced increase of conductance is

$$\Delta\sigma_{PCE} = q(\mu_n + \mu_p)p_{ph} + q\mu_p p_t, \quad (12)$$

which is enlarged by $q\mu_p p_t$ due to the presence of traps. As discussed above (see Figure 3c) we find that for $V_g \ll V_{th}$ the measured $I_{PC}$ is linear with the power density. This is compatible with a PCE of the form given in equation (12) under the reasonable assumption that $p_{ph} \gg p_t$ (further discussed in Supplementary Section S3).

Secondly, the trapped states $p_t$ partially screen the electric field arising from $V_g$, giving an additional contribution to the photocurrent due to the PGE. As we argue below, we believe that this contribution (not considered in earlier literature) is responsible for the fast-response PGE observed experimentally at high light-modulation frequencies.

Following equation (4), the shift in the threshold voltage due to the trapped charge carriers, $p_t$ is given by

$$\Delta V_{PGE} = -\left(\frac{1}{C_g} + \frac{1}{C_q}\right)ep_t = \frac{p_t}{\beta} \quad (13)$$

where $C_g$ is the geometrical capacitance, $C_q$ is the quantum capacitance, defined as $C_q = e^2 g_{2D}$ ($g_{2D}$ being the density of states of a two-dimensional electron gas) and $\beta = 7.17 \times 10^{17} \text{cm}^{-2}\text{V}^{-1}$ for our device (see Supplementary Section S4). The photocurrent produced by this voltage shift, $I_{PGE}$ is

$$I_{PGE} = \Delta V_{PGE}\frac{dI_{ds}}{dV_g} = \frac{p_t}{\beta}\frac{dI_{ds}}{dV_g}. \quad (14)$$

Finally, using equation (11) we obtain

$$I_{PGE} = \frac{D_t}{\beta}\frac{dI_{ds}}{dV_g}\frac{1}{1 + \frac{D_t}{\phi_A \tau_r}\left(\frac{\tau_t}{\tau_d}\right)} = A\frac{1}{1 + \frac{B}{P_D}}, \quad (15)$$

where we have defined the parameters A and B as

$$A = \frac{D_t}{\beta}\frac{dI_{ds}}{dV_g}; \quad B = \frac{D_t hc}{\eta\lambda\tau_r}\left(\frac{\tau_t}{\tau_d}\right), \quad (16)$$

We now use equation (15) to fit the measured power dependence for $V_g - V_{th} = 20$ V. Figure 4b shows the measured power dependence of $I_{PC}$ for $V_g - V_{th} = 20$ V and its fitting to equation (15), using A and B as fitting parameters. For comparison, we also show the best fit to the phenomenological equation $I_{PC} \propto P_D^\alpha$, commonly found in literature. While both fitting curves have a similar shape, our model allows us to better reproduce the experimental data points. For the norm of residuals ($r$) of the fittings to equation (15) we get $r_A = 10$ pA and $r_B = 7$ pA for excitons $X^A$ and $X^B$ respectively, roughly twice smaller than the values obtained for the fitting to $I_{PC} \propto P_D^\alpha$ ($r_A = 20$ pA and $r_B = 16$ pA). From the obtained fitting parameters A and B we can now extract an estimation for the density of trap states $D_t \approx 1 \times 10^{10}$ cm$^{-2}$, as well as the ratio of characteristic times $\tau_t(\tau_r\tau_d)^{-1} = 8.5 \times 10^3$ s$^{-1}$.

Finally, we estimate the energy level associated to the shallow hole traps, $E_T$, over the top of the valence band at $E_V$ by considering the detailed balance principle for the transitions between these states. Such condition for this particular case reads[6]

$$E_{T,V} = E_T - E_V = k_B T \ln\left(\frac{N_V \tau_d}{D_t \tau_t}\right). \quad (17)$$

Here $k_B$ is the Boltzmann constant and $N_V$ is the effective density of states of the valence band, given as $N_V = gm^* K_B T/(\pi\hbar^2)$. In our 2D system $g = 2$ due to the valley degeneracy and the efective mass of the carriers is $m^* = 0.4\, m_0$ with $m_0$ being the free electron mass.[22]

Last, since there are clear evidences that the recombination time $\tau_r$ is within the order of few picoseconds at low temperature,[23,24] we take the value of $\tau_r \approx 5$ ps to estimate the energy of the hole traps relative to the top of the valence band as $E_{T,V} \approx 8.4$ meV. This finding suggests the existence of shallow hole levels with energy very close to the valence band edge. As discussed below, we associate these levels with the presence of sulfur vacancies in the MoS$_2$ crystal.

## Conclusions

In all, we clearly identified two different regimes for photocurrent generation, that can be distinguished by their different dependence on the illumination power density $P_D$. For $V_g < V_{th}$, where the 1L-MoS$_2$ conduction band is fully depleted, $I_{PC}$ is linear with $P_D$, indicating that photocurrent is produced by PCE. In contrast, for $V_g > V_{th}$, there are three mechanism contributing to photoresponse: slow-response PGE (most likely due to polar molecules), a fast-response PGE (which we attribute to sulfur vacancies) and a PCE. In this latter case, the power dependence of the photocurrent becomes sublinear, indicating that the two PGE mechanisms are dominant over PCE.

In earlier works,[6] PGE in 1L-MoS$_2$ devices was attributed to a slow charge-trapping process by polar adsorbates in the vicinity of the 2D channel. However, here we find that the PGE dominates the photoresponse of the device even at frequencies as high as 1 KHz. We attribute this fast PGE to the effect of charge accumulation in shallow impurities near the 1L-MoS$_2$ valence band. By fitting the experimentally observed power dependence of $I_{PC}$ to a modified Hornbeck–Haynes model that includes this effect we can estimate the density of trap states to be $D_t \approx 1 \times 10^{10}$ cm$^{-2}$. Thus, even for relatively low trap densities, charge accumulation in shallow impurities can be the dominant mechanism for photoresponse.

The fitting mentioned above also allowed us to estimate the energy of the trap states to be of the order of 8 meV above the valence band edge. We considered different potential origins for these traps, including both native defects in the MoS$_2$[22] and extrinsic defects such as defects arising due to the h-BN encapsulation.[13,25] One of the most common defects in MoS$_2$, especially if it is fabricated by exfoliation, are sulfur vacancies. Ab-initio simulations of these defects[16–18] indicate that they support the existence of two families of states within the energy gap: a branch of states lying slightly above the middle of the gap, and a second branch lying very close to the valence band edge (which energy depends on the particular set of simulation parameters). Special attention to the latter branch has been paid in ref. [16], where the authors claim that these states present acceptor-like behaviour. Based on this evidence, we believe that the origin of the fast-responding PGE found in this work is related to the presence of sulfur vacancies in the 1L-MoS$_2$ channel.

## Experimental details

*Device fabrication and contact geometry* – We use a dry-transfer method based on the use of polypropylene carbonate (PPC) films[26] for fabricating the heterostructure of single layer (1L) MoS$_2$ completely encapsulated in hexagonal boron nitride (h-BN). The MoS$_2$ and h-BN flakes are first exfoliated by the standard scotch-tape method and transferred onto SiO$_2$/Si substrate. Then, we use optical microscopy to identify the 1L-MoS$_2$ flakes and confirm their thickness by micro-Raman spectroscopy (see Supplementary Section S5). We also select two h-BN flakes with thicknesses of 15-20 nm for the top layer h-BN and 25-30 nm for the bottom layer one (determined by their optical contrast).

Next, we transfer the top h-BN onto the MoS$_2$ flake and remove the remaining PPC by cleaning the sample with anisole, acetone and isopropanol (IPA) for few minutes. Both flakes are then picked up together with a PPC film and transferred onto the bottom h-BN. Finally, we perform a last cleaning with anisole, acetone and IPA, followed by an annealing in argon to remove any remaining PPC and bubbles in the heterostructure.[27]

The device geometry is defined by electron beam lithography (EBL) using PMMA as resist. For developing the resist we use a mixture of 1 part MIBK to 3 parts of isopropanol.[28] We etch away the EBL-exposed areas by dry plasma etching in a SF$_6$ atmosphere (40 sccm, P=75W, process pressure 6 mTorr and T= 10 ºC)[29]. The sides of the resulting etched structure have a pyramidal profile, necessary for a successful fabrication of edge contacts.

After defining the stack geometry, we fabricate the metallic contacts by a second EBL process followed by e-beam evaporation of 5 nm of titanium and 45 nm of gold. To prevent oxidation of the edge contacts all the fabrication steps described above are carried out in a single day. An optical image of the final device is presented in Supplementary Figure S5.

*Electrical and optoelectronic measurements* – The measurements are realized while keeping the sample inside a pulse-tube cryostat with an optical access. Drain-source and transfer *IV* characteristics are measured in two-terminal configuration using a two-channel sourcemeter unit (Keithley 2614B) The light source is a supercontinuum (white) laser (SuperK Compact), and the excitation wavelength is selected using a monochromator (Oriel MS257 with 1200 lines/mm diffraction grid). This allows to scan the visible and NIR spectral range, roughly from 450 nm to 840 nm. For AC optoelectronic measurements, the optical excitation is modulated by a mechanical chopper and the electrical response of the device is registered using a lock-in amplifier (Stanford Research SR830).

## Author Contributions

E. Diez and J.Q. conceived and supervised the research, D.V., Y.M.M., and J.Q developed and tested the experimental setup for photocurrent spectroscopy. V.C., J.S.-S. and J.Q. fabricated and characterized the monolayer MoS$_2$ phototransistor, D.V. and J.Q carried out the electronic, optoelectronic, and spectral measurements and data analysis, D.V., J.Q., E.Díaz, L.C. and F.D.-A. performed the theoretical analysis. The article was written through contribution of all the authors, coordinated by J.Q.

## Conflicts of interest

There are no conflicts to declare.

## Acknowledgements

The acknowledgements come at the end of an article after the conclusions and before the notes and references. We acknowledge financial support from the Agencia Estatal de Investigación of Spain (Grants PID2019-106820RB, RTI2018-097180-B-100, and PGC2018-097018-B-I00) and the Junta de Castilla y León (Grants SA256P18 and SA121P20), including funding by ERDF/FEDER. J.Q. acknowledges financial support from MICINN (Spain) through the programme Juan de la Cierva-Incorporación. We are also thankful to Mercedes Velázquez for her help with the photoluminescence and Raman characterization and to Adrián Martín-Ramos for his assistance on the development of the photocurrent measurement setup.

## Notes and references

# Supplementary Information to:
# Fast-response photogating in monolayer MoS$_2$ phototransistors

*Daniel Vaquero[1], Vito Clericò[1], Juan Salvador-Sánchez[1], Elena Díaz[2], Francisco Domínguez-Adame[2], Leonor Chico[2,3], Yahya M. Meziani[1], Enrique Diez[1] and Jorge Quereda[1]\**

[1] Nanotechnology Group, USAL–Nanolab, Universidad de Salamanca, E-37008 Salamanca, Spain

[2] GISC, Departamento de Física de Materiales, Universidad Complutense, E-28040 Madrid, Spain

[3] Instituto de Ciencia de Materiales de Madrid, CSIC, E-28049 Madrid, Spain

\* e-mail: j.quereda@usal.es

**Table of contents**

## S1. Room-temperature measurements

In this section we present photocurrent measurements at room temperature, analogous to the low-temperature measurements showed in the main text.

Supplementary Figure S1a shows a room temperature photocurrent spectrum acquired at $V_g - V_{th} = -10$ V with a power density of 1 mW cm$^{-2}$, as well as its fit to a quintuple Lorentzian, corresponding to the five main exciton transitions described in the main text ($T_A$, $X_{1s}^A$, $T_B$, $X_{1s}^B$ and $X_{2s}^A$). Due to the thermal energy, the peaks of the spectrum are broadened and red-shifted in comparison with the low temperature photocurrent spectrum presented in Supplementary Section 2.

Supplementary Figure S1b shows the transfer curve of the device at $V_{ds} = 10$ V. As expected, the increase in the current near the threshold voltage is here less abrupt than at low temperature. Supplementary Figures 1c and 1d show the photocurrent as a function of the gate voltage $V_g$ for illumination at E=1.87 eV and two different light-modulation frequencies: $f = 31.81$ Hz (c) and $f = 1$ kHz (d). Consistently with the results and the theoretical model presented in the main text, the photocurrent is strongly correlated with the transconductance of the device regardless of the modulation frequency. It is worth remarking that, for the room-temperature measurements presented here, the photocurrent measured at low modulation frequency is roughly 8 times larger than the one measured at 1 kHz, indicating that the effect of slow-responding traps due to polar adsorbates is much stronger at room temperature than at $T = 5$ K.

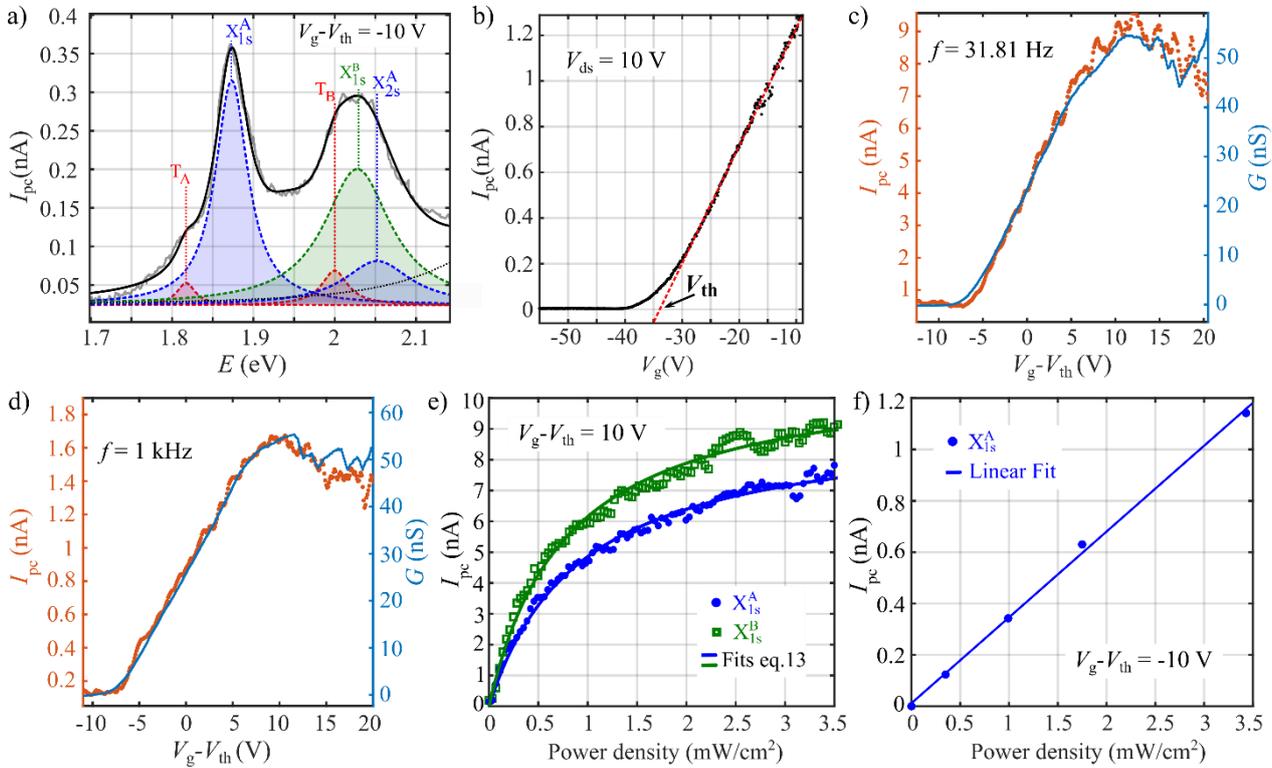

**Supplementary Figure S1**- Room temperature photocurrent measurements. (a) Photocurrent spectrum of the 1L-MoS2 phototransistor (gray solid line) and multi-lorentzian fitting (black solid line). The five main transitions are depicted in the figure. (b) Transfer curve of the device in dark at room temperature and $V_{ds}$=10V. (c-d) Gate dependence of the photocurrent depicted with the transconductance of the device in resonance with the exciton A at different frequencies of modulation (c) f = 31.81 Hz and (d) f = 1kHz. (e) Power dependence of the photocurrent at Vg-Vth=10V in resonance with the exciton A and B. The solid lines correspond to the fittings of the data to eq. 13. (f) Power dependence of the photocurrent at $V_g$-$V_{th}$ = − 10 V, in resonance with the exciton A.

Finally, Supplementary Figure S1e shows the power dependence of the photocurrent at $V_g - V_{th} = 10$ V, for two different illumination energies, matching the A and B exciton transitions, E=1.87 eV and E=2.01 eV respectively. The illumination power dependence of $I_{PC}$ is sublinear, as expected for the photogating effect. Similarly to our results at low temperature, the illumination power dependence of $I_{PC}$ becomes linear for gate voltages below the threshold voltage (see Supplementary Figure S1f), indicating that for this regime, the photoconductivity is dominated by the photoconductive effect.

## S2. Photocurrent spectroscopy

Supplementary Figure S2 shows a photoconductivity spectrum of our device acquired for $V_{ds} = 10$ V, and $V_g - V_{th} = -30$ V. At low temperature the main excitonic spectral features can be clearly resolved, with bandwidths as low as 8 meV.[1] The spectrum presents two main peaks corresponding to the A and B neutral excitons ($X_{1s}^A$ and $X_{1s}^B$ respectively), as well as three smaller features corresponding to the trion states ($T^A$ and $T^B$) and the 2s excited state $X_{2s}^A$ of the A exciton.

Detailed information on the experimental setup for photocurrent spectroscopy, as well as an in-depth analysis of the spectral features in 1L-MoS$_2$ transistors can be found in ref. [1]. The sample is placed inside a pulse-tube cryostat ($T$ = 5 K) and the whole device is exposed to laser illumination through an optical access. For illumination we use a SuperK Compact supercontinuum laser from NKT photonics, and the excitation wavelength is selected by an Oriel MS257 monochromator (1200 lines/mm). Using this light source allows us to scan the spectral range from 450 nm to 840 nm. The excitation signal is modulated by an optical chopper and the photocurrent is registered by a lock-in amplifier.

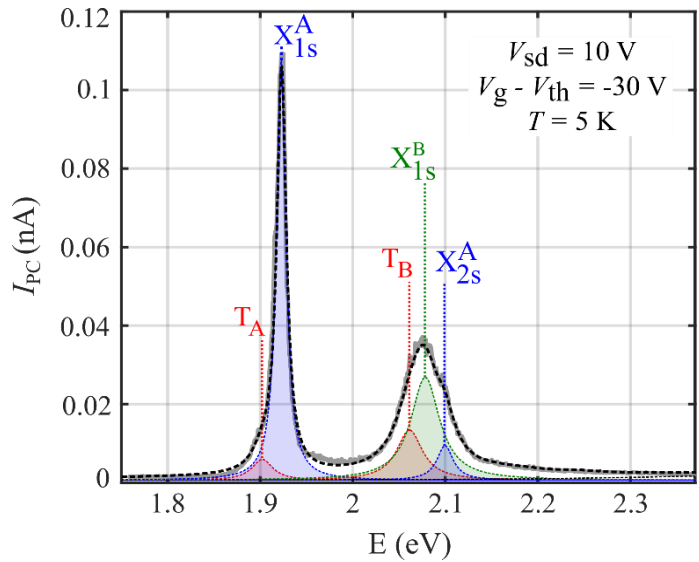

**Supplementary Figure S2.** Low-temperature photocurrent spectrum of the encapsulated 1L-MoS$_2$ device.

## S3. Extended discussion on power dependence of PCE

In the main text (equation 10) we obtained an expression for the increase of photoconductivity caused by the photoconductive effect (PCE):

$$\Delta\sigma_{\text{PCE}} = q(\mu_n + \mu_p)p_{\text{ph}} + q\mu_p p_t, \tag{18}$$

Replacing $p_{\text{ph}}$ and $p_t$ by their expressions (equations 8 and 9 in the main text) we get:

$$\Delta\sigma_{\text{PCE}} = q\tau_r(\mu_n + \mu_p)\phi_A + q\mu_p \frac{\phi_A D_t \tau_r}{\phi_A \tau_r + D_t\left(\frac{\tau_t}{\tau_d}\right)}, \tag{19}$$

The first term in the right-hand side of equation (19) is linear with the power (note that $\phi_A \propto P_D$). Thus, in absence of trap states, *i.e.* for $D_t = 0$, $\Delta\sigma_{\text{PCE}}$ is linear with the power (as long as the main relaxation mechanism is Shockley-Read-Hall recombination). When a finite density of traps is present, it is useful to consider the three following situations:

(i) $\quad \phi_A \gg D_t \frac{\tau_t}{\tau_r \tau_d}$

This is the relevant scenario illumination power densities large enough for the trap states to become saturated. Under this situation, the second right-hand term in equation (19) can be simplified as

$$\frac{q\mu_p \phi_A D_t \tau_r}{\phi_A \tau_r + D_t\left(\frac{\tau_t}{\tau_d}\right)} \approx q\mu_p D_t, \tag{20}$$

which gives only a constant contribution to $\Delta\sigma_{\text{PCE}}$. Thus, the total photoconductivity remains linear with the power:

$$\Delta\sigma_{\text{PCE}} \approx q\tau_t(\mu_n + \mu_p)\phi_A + q\mu_p D_t, \tag{21}$$

(ii) $\quad \phi_A \ll D_t \frac{\tau_t}{\tau_r \tau_d}$

For very low illumination power, the density of available trap states only changes by a very small amount due to light exposure. Under this situation, the second right-hand term in equation (19) can be again simplified as

$$\frac{q\mu_p \phi_A D_t \tau_r}{\phi_A \tau_r + D_t\left(\frac{\tau_t}{\tau_d}\right)} \approx q\mu_p \frac{\tau_d \tau_r}{\tau_t} \phi_A, \tag{22}$$

which now gives a linear contribution to $\Delta\sigma_{\text{PCE}}$. Again, the total photoconductivity remains linear with the power:

$$\Delta\sigma_{\text{PCE}} \approx \left(q\tau_t(\mu_n + \mu_p) + q\mu_p \frac{\tau_d \tau_r}{\tau_t}\right)\phi_A, \tag{23}$$

In this scenario, the effect of localized states is to enhance the slope of $\Delta\sigma_{\text{PCE}}$ while keeping it linear with the power density.

(iii) $\quad \phi_A \approx D_t \frac{\tau_t}{\tau_r \tau_d}$

Finally, for intermediate power densities, equation (19) cannot be simplified and the presence of localized states results in a sublinear contribution to photocurrent.

## S4. Estimation of carrier density and Fermi energy shift

*Note: The measurements presented in this article were performed in the same device studied in an earlier publication by the authors[1]. This supplementary section is reprinted from the Supplementary Information of the mentioned publication for convenience of the readers.*

In the following we use a capacitor model to estimate the increase in carrier density $\delta n$ produced by the gate voltage. The gate voltage $V_g$, *i.e.* the total voltage drop between the Si back gate and the MoS$_2$ channel, will be given by

$$\delta V_g = \delta E \cdot d + \frac{1}{e} \delta E_F \qquad (24)$$

Where $E$ is the electric field between the electrode and the flake, $-e$ is the electron charge and $E_F$ is the Fermi energy. For a parallel plate with two different insulator layers the geometrical capacitance is

$$C_g = \left( \frac{d_{SiO_2}}{\epsilon_0 \epsilon_{SiO_2}} + \frac{d_{BN}}{\epsilon_0 \epsilon_{BN}} \right)^{-1}, \qquad (25)$$

and we have

$$\delta E \cdot d = \frac{e \delta n}{C_g}. \qquad (26)$$

Replacing in (S8) and using $\delta E_F = (\delta E_F/\delta n)\, \delta n = \delta n/D$, where $D$ is the density of states of the 2D semiconductor, we get

$$\delta V_g = \frac{de}{\epsilon_0 \epsilon_d} \cdot \delta n + \frac{1}{eD} \delta n = \left( \frac{1}{C_g} + \frac{1}{C_q} \right) e \delta n, \qquad (27)$$

where we have defined the quantum capacitance as $C_q = e^2 D$. We can now express equation (27) in terms of the Fermi energy using $\delta E_F = \delta n/D$. We get

$$\delta V_g = \left( \frac{1}{C_g} + \frac{1}{C_q} \right) eD\, \delta E_F = \left( \frac{1}{C_g} + \frac{1}{C_q} \right) \frac{C_q}{e} \delta E_F. \qquad (28)$$

Therefore, solving for $E_F$, we have

$$\delta E_F = \frac{e \delta V_g}{1 + \dfrac{C_q}{C_g}} \ . \tag{29}$$

We model the density of states of 1L-MoSe$_2$ as the step function

$$D(E) = \begin{cases} g_{2D} \equiv \dfrac{\mu_{\text{eff}}}{\pi \hbar^2} & \text{if } E > E_{\text{CB}} \\ 0 & \text{if } E > E_{\text{CB}} \end{cases}, \tag{30}$$

where $\mu_{\text{eff}}$ is the electron effective mass in MoS$_2$ ($\mu_{\text{eff}} = 0.35\, m_0$) and $E_{\text{CB}}$ is the edge of the conduction band. Then, by integrating equation (29) we get

$$\Delta E_F = \frac{e}{1 + \dfrac{e^2 g_{2D}}{C_g}} (V_g - V_{\text{th}}) \ , \tag{31}$$

where $V_{\text{th}}$ is the threshold voltage at which $E_F = E_{\text{CB}}$. In our case, we get $\Delta E_F/(V_g - V_{\text{th}}) = 0.28$ meV V$^{-1}$, which for the maximal gate voltages applied here ($V_g - V_{\text{th}} = 50$V) gives $\Delta E_F = 14$ meV. Finally, the density of excess carriers, $n$ can be obtained as $n = \Delta E_F \cdot g_{2D} = 7.17 \times 10^{10} \text{cm}^{-2} \text{V}^{-1} (V_g - V_{\text{th}})$. Thus, the maximal carrier densities reached here are of $n = 3.58 \times 10^{12} \text{cm}^{-2}$.

## S5. Raman and photoluminescence characterization

*Note: The measurements presented in this article were performed in the same device studied in an earlier publication by the authors[1]. This supplementary section is reprinted from the Supplementary Information of the mentioned publication (with minor changes) for convenience of the readers.*

We determine the thickness of the MoS$_2$ flakes used for device fabrication by a combination of optical microscopy, Raman mapping and photoluminescence. Supplementary Figure S3a shows an optical microscope image of the MoS$_2$ flake used to fabricate the device described in the main text, and Supplementary Figure S3b shows a false color map of the ratio between the summed intensities of the A$_{1g}$ + E$^1_{2g}$ Raman peaks of MoS$_2$ and the intensity of the Si peak, in logarithmic scale. The different thicknesses can be clearly distinguished in the figure. Supplementary Figure S3c shows individual spectra acquired at the different regions labelled in Supplementary Figure S3a. The number of layers can be here confirmed by the difference between the spectral positions of the E$^1_{2g}$ and A$_{1g}$ peaks, $\Delta f$.[2,3] For the thinnest region we obtain $\Delta f = 19.4$ cm$^{-1}$, compatible with the values given in literature for 1L-MoS$_2$.

We further confirm the thickness of the MoS$_2$ flakes by measuring the position of the A exciton peak in their photoluminescence spectrum. Supplementary Figure S4 shows a room-temperature photoluminescence spectrum acquired at the monolayer MoS$_2$ flake. The $X^A_{1s}$ exciton peak can be clearly observed at around 1.87 eV, in good agreement with the values found in literature.[3–5]

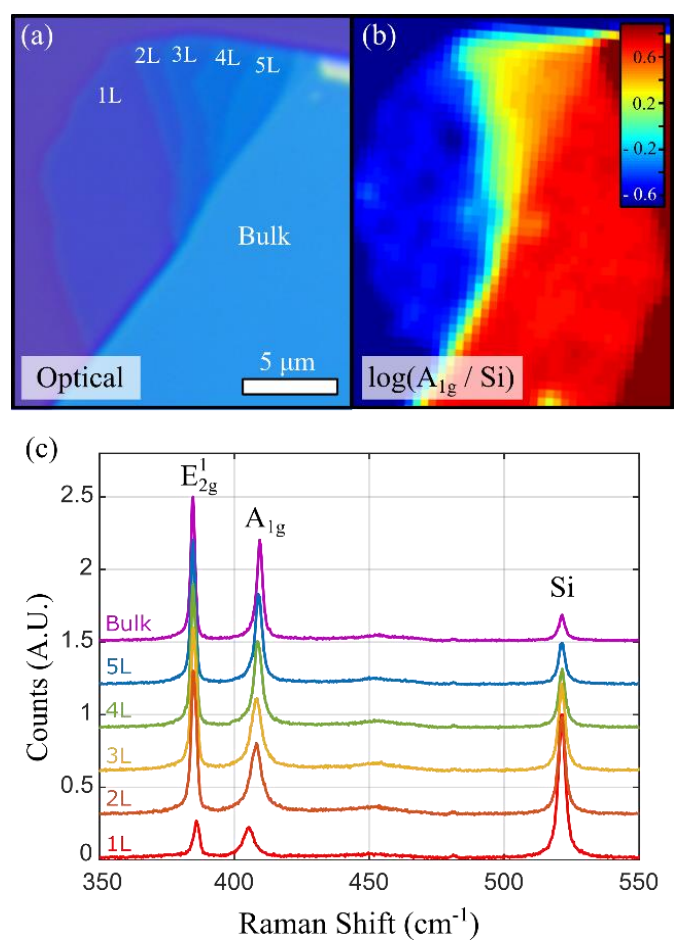

**Supplementary Figure S3.** Raman characterization of the MoS$_2$ thickness. (a) Optical microscopy image of the MoS$_2$ flake used for the device of the main text. The labels indicate regions with different thickness. (b) False color Raman map of the difference between the A$_{1g}$ and Si peak intensities, as labeled in panel c. (c) Raman spectra acquired at the different regions labelled in Figure 1a. The spectra show three prominent peaks corresponding to the A$_{1g}$ and E$^1_{2g}$ Raman modes of MoS$_2$ and the Si Raman mode.

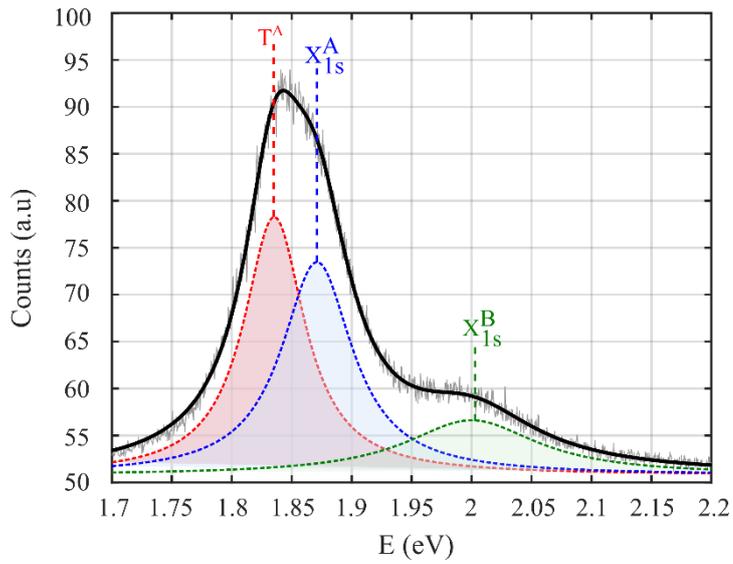

**Supplementary Figure S4.** Room-temperature photoluminescence spectra of monolayer $MoS_2$ on $SiO_2$ under 530 nm excitation. The dashed lines are the individual contributions from the TA, $X_{1s}^A$ and $X_{1s}^B$ exciton transitions.

## S6. Optical image of the device

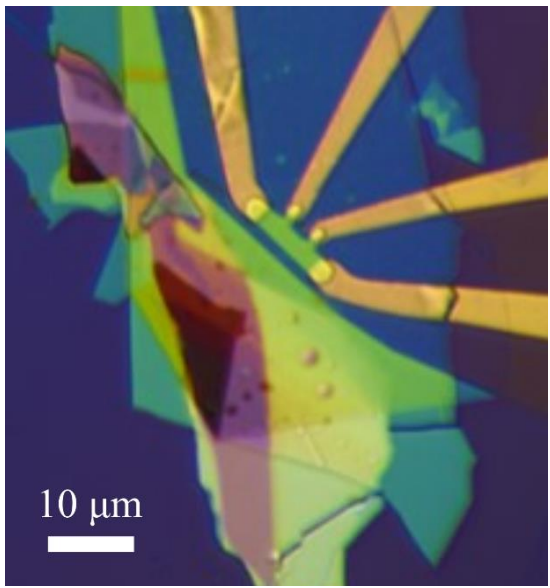

**Supplementary Figure S5.** Optical image of the encapsulated 1L-$MoS_2$

**Supplementary Note 7. Derivation of photoconductive gain**

It can be convenient in some cases to express the device's photoresponse in terms of the photoconductive gain $G_{ph}$. This quantity is defined as the ratio between the number of charge carriers collected by the electrodes and the number of absorbed photons:

$$G_{ph} = \frac{\text{collected charge carriers}}{\text{absorbed photons}} = \frac{q^{-1}I_{\text{PC}}}{\phi_A W L} \tag{325}$$

where $q$ is the elementary charge, and W and L are the width and length of the semiconductor channel, respectively.

In a trap-free semiconductor, in absence of photogating effect, the photoconductive gain can be obtained as [6]

$$G_{ph} = \frac{\tau_r}{\tau_{\text{tr,n}}} + \frac{\tau_r}{\tau_{\text{tr,p}}} \tag{336}$$

where $\tau_r$ is the electron-hole recombination lifetime and $\tau_{\text{tr,n}}$ ($\tau_{\text{tr,p}}$) is the transit time for electrons (holes), *i.e.* the time required for an electron (hole) to drift across the semiconductor channel, from the source to the drain electrode.

Let us now derivate the expression of $G_{ph}$ in the presence of shallow states such as the ones considered in the main text. To do so, it results convenient to separate $I_{\text{PC}}$ into its photoconductive ($\Delta I_{\text{PCE}}$) and photogating ($\Delta I_{\text{PGE}}$) contributions. For $\Delta I_{\text{PCE}}$, combining equations 2 and 3 of the main text we have

$$\Delta I_{\text{PCE}} = \frac{W}{L} V_{\text{ds}} q (\mu_n n_{\text{ph}} + \mu_p p_{\text{ph}}). \tag{17}$$

Assuming a uniform electric field $E$ across the channel, we can write $V_{\text{ds}} = EL$. Then, reordering terms we have

$$\Delta I_{\text{PCE}} = qWL \left( \frac{E\mu_n}{L} n_{\text{ph}} + \frac{E\mu_p}{L} p_{\text{ph}} \right) = qWL \left( \frac{E(\mu_n + \mu_p)}{L} p_{\text{ph}} + \frac{E\mu_p}{L} p_t \right). \tag{18}$$

Equation 18 can now be rewritten in terms of the electron and hole transit times $\tau_{\text{tr,n}} = L/E\mu_n$ and $\tau_{\text{tr,p}} = L/E\mu_p$. This yields

$$\Delta I_{\text{PCE}} = qWL \left( p_{\text{ph}} \left( \frac{1}{\tau_{\text{tr,n}}} + \frac{1}{\tau_{\text{tr,p}}} \right) + \frac{p_t}{\tau_{\text{tr,p}}} \right). \tag{19}$$

We now replace $p_{\text{ph}}$ and $p_t$ by their expressions from equations 9 and 10 of the main text:

$$\Delta I_{\text{PCE}} = qWL\left(\phi_A \tau_r\left(\frac{1}{\tau_{\text{tr,n}}} + \frac{1}{\tau_{\text{tr,p}}}\right) + \frac{\tau_r}{\tau_{\text{tr,p}}} \frac{\phi_A D_t}{\phi_A \tau_r + D_t\left(\frac{\tau_t}{\tau_d}\right)}\right). \tag{19}$$

Finally, using equation 15 for the photoconductive gain we get

$$G_{\text{ph,PCE}} = \frac{q^{-1}\Delta I_{\text{PCE}}}{\phi_A WL} = \frac{\tau_r}{\tau_{\text{tr,n}}} + \frac{\tau_r}{\tau_{\text{tr,p}}} + \frac{\tau_r}{\tau_{\text{tr,p}}} \frac{\phi_A D_t}{\phi_A \tau_r + D_t\left(\frac{\tau_t}{\tau_d}\right)}. \tag{20}$$

For $\Delta I_{\text{PGE}}$ equation 14 in the main text gives

$$I_{\text{PGE}} = \frac{D_t}{\beta} \frac{dI_{ds}}{dV_g} \frac{1}{1 + \frac{D_t}{\phi_A \tau_r}\left(\frac{\tau_t}{\tau_d}\right)}, \tag{21}$$

Which corresponds to a photoconductive gain of

$$G_{\text{ph,PGE}} = \frac{D_t}{qWL\beta} \frac{dI_{ds}}{dV_g} \frac{1}{\phi_A + \frac{D_t}{\tau_r}\left(\frac{\tau_t}{\tau_d}\right)}, \tag{22}$$

The total photoconductive gain in the device will be the sum of the two contributions:

$$G_{\text{ph}} = G_{\text{ph,PCE}} + G_{\text{ph,PGE}} \tag{23}$$

$$G_{\text{ph}} = \frac{\tau_r}{\tau_{\text{tr,n}}} + \frac{\tau_r}{\tau_{\text{tr,p}}} + \left(\frac{\phi_A \tau_r}{\tau_{\text{tr,p}}} + \tau_r \frac{1}{qWL\beta} \frac{dI_{ds}}{dV_g}\right) \frac{D_t}{\phi_A \tau_r + D_t\left(\frac{\tau_t}{\tau_d}\right)}, \tag{24}$$

Thus, the presence of shallow traps results in an increase in the photoconductive gain, compared to the trap-free situation. Note that, if the density of trap states is set to zero, $D_t = 0$, the third term in the right-hand side of equation 24 cancels out, and we recover the expression of $G_{\text{ph}}$ for a trap-free scenario (equation 15).